\documentclass[12pt,a4paper]{article}
\begin{document}
\sloppy
\title{\bf{Signatures of TeV Scale Gravity in High Energy
Collisions}}
\author {A. Nicolaidis \\ Department of Theorical Physics \\ University of
Thessaloniki \\ 54124 Thessaloniki, Greece \\ nicolaid@auth.gr \\ \\
N.G. Sanchez \\ Observatoire de Paris LERMA, UMR 8540, CNRS \\61, Avenue de
l'Observatoire \\ 75014 Paris, France \\ Norma.Sanchez@obspm.fr}
\date{ }
\maketitle

\begin{center}
{\bf Abstract}
\end{center}
In TeV scale unification models, gravity propagates in 4+$\delta $
dimensions while gauge and matter fields are confined to a four
dimensional brane, with gravity becoming strong at the TeV scale. For a
such scenario, we study strong gravitational interactions in a effective
Schwarzschild geometry. Two distinct regimes appear. For large
impact parameters, the ratio $\rho \sim (R_s/r_0)^{1+\delta }$, (with
$R_s$ the Schwarzschild radius and $r_0$ the closest approach to the black
hole), is small and the deflection angle $\chi $ is proportional to $\rho $
(this is like Rutherford-type scattering). For small impact parameters, the
deflection angle $\chi $ develops a logarithmic singularity and becomes
infinite for $\rho = \rho _{crit} = 2/(3+\delta )$. This singularity is
reflected into a strong enhancement of the backward scattering (like a
glory-type effect). We suggest as distinctive signature of black hole
formation in particle collisions at TeV energies, the observation of
backward scattering events and their associated diffractive
effects. \\
\newpage
The main motivation for introducing new physics comes from the need to
provide a unified theory in which two disparate scales, ie the electroweak
scale $M_W \sim $ 100 GeV and the Planck scale $M_P \sim 10^{19}$ GeV, can
coexist (hierarchy problem). A novel approach has been proposed for
resolving the hierarchy problem \cite{a1}.
Specifically, it has been suggested that our four dimensional world is
embedded in a higher dimensional space with D dimensions, of which
$\delta $ dimensions are compactified with a relatively large (of order of
mm) radius. While the Standard Model (SM) fields live on the 4-dimensional
world (brane), the graviton can propagate freely in the higher dimensional
space (bulk).\\
The fundamental scale $M_f$ of gravity in D dimensions is related to the
observed 4-dimensional Newton constant $G_N$ by
\begin{equation}
G_N=\frac{1}{V_{\delta }}\: \left( \frac{1}{M_f}\right) ^{(2+\delta )}
\end{equation}
where $V_{\delta }$ is the volume of the extra dimensional space. A
sufficiently large $V_{\delta }$ can then reduce the fundamental scale of
gravity $M_f$ to TeV energies , which is not too different from $M_W$,
thereby resolving the hierarchy problem. \\ \\
The prospect of gravity becoming strong at TeV energies, opens the
possibility of studying gravity in particle collisions at accessible
energies (at present or in the near future). To that respect, salient
features of the cosmic ray spectrum (the ''knee'') have been attributed to
gravitational bremsstrahlung \cite{kn}. By reproducing the cosmic ray
spectrum, the parameters of the low scale gravity can be inferred
($\delta \sim 4$ and $M_f \sim $ 8 TeV).\\
In this letter, we would like to study the interactions among particles,
mediated by low scale gravity. The optimum would be to address this issue
within a complete quantum gravity theory. But we are lacking this
framework.
We could turn to pertubative quantum gravity. But we would miss all the
essential features of strong gravity we are interested in.
Furthermore, perturbative calculations are uncertain, since there is no
definite method for summing up the Kaluza-Klein contributions \cite{gs}.
We prefer to work within an effective  approach in which gravity
effects are treated classically but non-perturbatively. That is, the effect
of particle collisions at TeV energy scale is considered as the scattering
of a particle in the effective {\it curved} background produced by all the
others. While our estimates would not be the exact answer, we anticipate
that they would reflect the main characteristics of strong gravity
effects.
Several arguments support this approach. Particle collisions in
string theory are, within the eikonal approximation, like the classical scattering of a particle
in the  effective gravitational shock wave background created by all
the others \cite{la},\cite{lb}. For large impact
parameters, the shock wave profile is
of Aichelburg-Sexl type (point particle source). For intermediate
impact parameters, the shock wave profile is different, corresponding
to a localized extended source \cite{lc}. The shock wave background description is only
valid for large or intermediate impact parameters, namely weak gravity
limit. The scattering phase shift in the Aichelburg-Sexl background
just reproduces the phase shift of the newtonian gravitational tail.
For small impact parameters (strong gravity effects)
 a full black hole background is necessary.
\\ \\
We consider a SM particle moving under the influence of a (4+$\delta $)-
dimensional black-hole. The effective metric is \cite{mp}
\begin{equation}
ds^2=f(r)dt^2-\frac{1}{f(r)}\, dr^2-r^2(d\theta ^2+\sin ^2\theta d\varphi
^2)
\end{equation}
where
\begin{equation}
f(r)=1-\left( \frac{R_s}{r}\right) ^{1+\delta }
\end{equation}
\begin{equation}
R_s=\frac{1}{\sqrt{\pi }M_f}\left[ \frac{M}{M_f}\:  \frac{8\Gamma \left(
\frac{\delta +3}{2}\right)} {\delta +2}\right] ^{1/(\delta +1)}
\end{equation}
$M_f \simeq $ TeV is related to $G_N$ by eq (1). For particle collisions,
$M=\sqrt{s}$, the center-of-mass energy. A particle impinging upon the
black
hole at impact parameter b will approach the black hole at a closest
distance
$r_0$, related by
\begin{equation}
b^2=\frac{r_0^2}{1-\rho }
\end{equation}
with
\begin{equation}
\rho =\left( \frac{R_s}{r_0}\right) ^{1+\delta }
\end{equation}
The deflection angle $\chi $ of the particle is given by (details will be
reported elsewhere)
\begin{equation}
\chi = 2\Phi _0 - \pi
\end{equation}
\begin{equation}
\Phi _0 = \int _0^1\frac{d\omega}{[1-\omega^2-\rho(1-\omega^{3+\delta})]^{1/2}}
\end{equation}
For large impact parameters, $\rho$ acquires small values. A Taylor
expansion
in $\rho$ provides then
\begin{equation}
\chi = I(\delta )\rho
\end{equation}
where $I(\delta)$ is a constant depending solely on $\delta$. Our result is
in agreement with the results obtained within pertubative (classical or
quantum)
gravity. Furthermore, by setting $\delta$ = 0 we find the traditional
Rutherford formula, or the equivalent formula for pertubative Yang-Mills
scattering in 4 dimensions. \\
Small b values give rise to larger values of $\rho$. We find that with
decreasing b, $\chi$ increases and there is a critical $\rho$  value where
$\chi$ becomes infinite. A reasonable approximation of $\chi$ for all
$\rho$
values is the following
\begin{equation}
\chi = a\, \ln \left( \frac{1}{1-\rho /\rho _{crit}}\right)
\end{equation}
with
\begin{equation}
\rho _{crit}=\frac{2}{(3+\delta )}
\end{equation}
\begin{equation}
a = \rho _{crit}\, I(\delta )
\end{equation}
The parameter $\rho _{crit}$ has a clear physical meaning. It corresponds
to
the unstable circular orbit around the black hole. As $\rho$ approachs
$\rho _{crit}$, the particle starts orbiting around the black hole before
escaping to infinity. At $\rho = \rho _{crit}$ the particle stays in
circular
orbit, implying an infinite value for $\chi$. For $\rho < \rho _{crit}$ the
particle is fully absorbed by the black hole. \\
Our findings, transformed into differential cross-section, provide
\begin{equation}
\frac{d\sigma _0}{d\Omega }\, (\chi )=CR^2_s\, \frac{1}{\sin \chi }\,
\frac{\exp (-\chi /a)}{[1-\exp (-\chi /a)]^\frac{(3+\delta )}{(1-\delta )}}
\end{equation}
where C is a $\delta$-dependent constant. Notice that the one-to-one
correspondence between the impact parameter b and the deflection angle
$\chi$ is lost
\begin{equation}
\rho=\rho _{crit}-\rho _{crit}\: e^{-\chi /a}
\end{equation}
Due to orbiting, different values of b give rise to the same $\chi$. We
have to sum over all $\chi _n=\chi + 2n\pi $ and the differential cross
section becomes
\begin{equation}
\frac{d\sigma (\chi )}{d\Omega }=\sum\limits^\infty _{n=0}\: \frac{d\sigma
_0}
{d\Omega }\: (\chi _n)
\end{equation}
At small $\chi$, the cross-section diverges like $(1/\chi )^{\gamma }$ with
$\gamma = (4+2\delta )/(1+\delta )$. (for $\delta =0$, $\gamma =4$). This
is
the well known focusing of forward scattering at $\chi$ = 0 due to the
large
range (newtonian or coulombian) interaction at large distances. At large
angles, the cross-section, although is relatively suppressed, it diverges
again  at $\chi = \pi $, like $\frac{1}{(\chi -\pi )}$ for any $\delta$.
This focusing at backward scattering is due to the strong attractive
black hole potential at short distances (it is not present in less
attractive gravitational fields, nor in Rutherford scattering).\\ \\
Imagine then collisions of cosmic rays particles in the atmosphere. At very
high energies, approaching the fundamental scale of gravity $M_f$,
gravitational interactions become important and it is expected that in
some events, backward scattering will occur. Experimental devices like
EUSO \cite{s}, OWL \cite{k}, will register the development of a shower
changing direction suddenly. We suggest that this type of events should be
seen as evidence for black hole formation in cosmic rays 
interactions, mediated by
TeV scale gravity. Similar situations may arise in detecting high
energy neutrinos by a neutrino telescope \cite{a2}. The induced energetic
muon might spiral and the emitted Cherenkov light will form a luminous halo
rather
than a forward cone. Again, these events should be classified as black hole
formation. \\
  Our calculation reveals another generic feature known as duality \cite{v}.
Gravity at large distances behaves like pertubative Yang-Mills fields at
short distances, both providing power-law behaviours. On the other hand,
gravity at short distances is described by exponential  cutoffs, very
similar to soft QCD phenomena at large distances. Clearly, this issue
desserves further study. \\ \\
There is an extensive literature on black hole formation in particle
collisions, within the framework of low scale gravity \cite{dl},\cite{gt}.
In
these  works, the detection of the emitted Hawking radiation has been
proposed as signal for black hole production. Hawking radiation
corresponds to particles just escaping the horizon and in a realistic
situation one should include the radiation emitted in the whole scattering
process. Furthermore a black-body spectrum for emitted particles is not
synonymous of Hawking radiation. In hadronic collisions at high energies
the
emitted particles follow  thermal spectra, without even implying
thermal equilibrium \cite{cg}. The relevant calculations 
\cite{dl},\cite{gt}, presssupose also
the validity of the parton model. However the formation of a black 
hole corresponds
to the strong gravity regime, with multiple gravitons being 
exchanged, and the hypothesis
of individual "free" partons is not justified.\\ \\
We anticipate that quantum effects will make milder the
$\chi = \pi$ singularity. In analogy to the wave scattering by a black hole
\cite{s2},
\cite{s3}, \cite{fhm}, diffractive scattering effects should be expected,
and interference
between the waves scattered at angles differing in 2n$\pi$, (n=1, 2,...)
will occur. A pattern of bright and dark rings (equivalent to the ``glory''
effect rings) will emerge,  a  distinctive feature of the presence of 
a black hole state formed in
the collisions. The S-matrix for such collisions should show an
oscillatory picture as a function of the scattering angle, with specific
location of peaks and dips between the two enhancements at $\chi = 0$ and
$\chi = \pi$. \\

Altogether we analysed gravitational scattering within the TeV-scale
gravity
models. Our treatment is classical but incorporates all non-perturbative
features of strong gravity. We find out that scattering mediated by
gravitational interaction develops a singularity at scattering angle
$\chi = \pi $. Only a black hole, a compact object associated with a
horizon can produce this effect.
We suggest then the observation of backward scattering events, associated
with rainbow-like diffraction patterns, as signals for the formation of black
hole
states in high energy particle collisions.
Experiments to be carried out in the near future would test the reality of
gravity becoming strong at TeV energies.

\end{document}